\documentstyle[a4,12pt,axodraw]{article}
\newcommand{\be}{\begin{equation}}
\newcommand{\nn}{\nonumber}
\newcommand{\bea}{\begin{eqnarray}}
\newcommand{\eea}{\end{eqnarray}}
\newcommand{\ba}{\begin{array}}
\newcommand{\ea}{\end{array}}
\newcommand{\ee}{\end{equation}}

\expandafter\ifx\csname mathbbm\endcsname\relax

\else

\fi \textheight 22cm \textwidth 15cm \topmargin 1mm \oddsidemargin
5mm \evensidemargin 5mm

\newcommand{\beas}{\begin{eqnarray*}}
\newcommand{\eeas}{\end{eqnarray*}}
\newcommand{\bes}{\begin{equation*}}
\newcommand{\ees}{\end{equation*}}
\newcommand{\dir}{\not\!\!{D}}

\newcommand{\lf}{\left}
\newcommand{\ri}{\right}
\newcommand{\f}{\frac}

\def\tr           {\mbox{\rm tr}\,}

\def\i2           {\mbox{$\frac{i}{2}$}}

\def\al           {\alpha}
\def\ald           {{\dot {\alpha}}}
\def\alb           {{\bar{\alpha}}}
\def\bet           {\beta}

\def\del           {\delta}

\def\ep           {\epsilon}

\def\ga           {\gamma}

\def\la           {\lambda}

\def\si           {\sigma}

\def\th{\theta}

\def\ran          {\rangle}
\def\lan          {\langle}

\def\alb           {{\bar{\alpha}}}

\def\osi{{\bar {\si}}}

\def\thb        {{\bar {\theta}}}

\def\ola        {{\overline \la}}
\def\ow         {{\overline W}}

\begin{document}

\begin{titlepage}
\hfill \vbox{
    \halign{#\hfil         \cr
           hep-th/0311137 \cr
           IPM/P-2003/075 \cr
           } 
      }  
\vspace*{20mm}
\begin{center}
{\LARGE {Comments on Gluino Condensates in ${\cal N}=1/2$ SYM Theory }\\
}

\vspace*{15mm} \vspace*{1mm} {Ali Imaanpur }

\vspace*{1cm}

{\it  Department of Physics, School of Sciences \\
Tarbiat Modares University, P.O. Box 14155-4838, Tehran, Iran\\
\vspace*{1mm}
Institute for Studies in Theoretical Physics and Mathematics (IPM)\\
P.O. Box 19395-5531, Tehran, Iran}\\
\vspace*{1cm}

\end{center}


\begin{abstract}
Using Ward identities of ${\cal N}=1/2$ supersymmetric Yang-Mills theory, we show 
that while the partition function and antichiral gluino condensates remain invariant 
under the $C$ deformation, chiral gluino correlators can get contributions from all 
gauge fields with instanton numbers $k\leq 1$. In particular, a Ward identity of the $U(1)_R$ symmetry allows us to determine the explicit dependence of chiral gluino correlators on the deformation parameter.

\end{abstract}

\end{titlepage}

\section{Introduction}
Four dimensional ${\cal N}=1$ supersymmetric gauge theories naturally arise 
as low energy effective field theories of D-branes when string theory is compactified 
over Calabi-Yau 3-folds. One could also turn on a graviphoton 
background, and look at the consequence it has on the effective theory of D-brane. 
The interesting observation is that in this background the odd coordinates 
of superspace become nonanticommuting over the brane \cite{GRASSI, CVAFA, KLEMM, SEI2}.     
This has the effect of reducing the number of supersymmetries to half. However, 
one can still construct a super Yang-Mills theory describing the effective theory on 
the brane \cite{SEI}. Different aspects of this model has been further studied in 
\cite{REY1}-\cite{GHOD}. 
While the generalization to 
${\cal N}=2$, along with other interesting features of noncommutative superspace
have further been explored in \cite{ARAKI}-\cite{ISO}.

In this note we are to study the effect of $C$ deformation on the 
gluino correlators. Using the Ward identity of the unbroken supersymmetry 
$Q$, we will see that the partition function as well as the antichiral gluino 
condensates remain invariant under the $C$ deformation. Namely they will have 
the same value as in ${\cal N}=1$ SYM $U(N)$ gauge theory, and in particular  
$\lan \tr (\ola\ola)^N \ran$ gets contribution only from 
one instanton sector. In contrast, chiral correlation functions of gluini  
can get contributions not only from one anti-instantons $(F^-=0)$, 
the only configurations which contribute to the chiral gluino correlators in 
${\cal N}=1$ theory, but also from gauge fields of topological charges $k\leq 0$. 
In what follows, we will discuss the nature of these field configurations, 
and their possible contributions to the chiral correlators using Ward identities 
of the anomalous $U(1)_R$ symmetry. Although $R$ symmetry is explicitly broken at the 
classical level, treating the deformation parameter $C$ as a background field with 
an $R$ charge the corresponding Ward identity tells us about the dependence 
of the correlators on the deformation parameter.  

As (anti)instantons play an important role in our discussion of (anti)chiral
correlators of gluini, let us briefly discuss how they 
might get deformed in the presence of fermions \cite{SOH}. 
In ${\cal N}=1/2$ SYM model, the equation of motion for the gauge field read
\be
D_\mu \lf ( F^{\mu\nu} + iC^{\mu\nu}\ola\ola \ri) = \si^\nu_{\al\ald}
\, \{\la^\al \, , \ola^\ald \} \, ,\label{F1}
\ee
while for  $\ola$ and $\la$ we have
\bea
&& \overline {\dir} \la =-C^{\mu\nu} F_{\mu\nu}^+\ola- i\f{|C|^2}{2}(\ola\ola)\ola \nn \\
&& \dir \ola =0 \, .\label{F2}
\eea
Note that when $F^-_{\mu\nu}=0$, the last equation has no solution. Setting $\ola=0$, 
then $F^-_{\mu\nu}=0$ solves the first equation, while the second equation reduces 
to the ordinary zero mode equation for chiral fermions. Therefore anti-instantons 
($F^-=0$) and the corresponding chiral zero modes are a solution to the above field 
equations. What about instantons ($F^+=0$)? In this background there are antichiral 
zero modes $\ola$, and hence the second equation implies that chiral fermions $\la$ 
cannot be zero either. On the other hand, using the Bianchi identity in the first equation 
we have 
\be
iC^{\mu\nu}D_\mu \lf (\ola\ola \ri) = \si^\nu_{\al\ald}
\, \{\la^\al \, , \ola^\ald \} \, .
\ee
Acting now with the covariant derivative $D_\nu$ 
on the above equation, and using $F^+=0$ implies that ${\overline {\dir}} \la=0$, 
which is inconsistent with the first equation of (\ref{F2}). So we conclude that 
instantons cannot be a solution to the equations ({\ref{F1}) and ({\ref{F2}). 
In \cite{SOH} it was argued that 
instanton like solutions are still possible if we instead consider the deformed instanton 
equations
\bea
&& F_{\mu\nu}^+ +\f{i}{2}C_{\mu\nu}\ola\ola =0 \nn \\
&& \dir \ola =0 \, \nn \\
&& {\overline {\dir}} \la =0 \, , \label{BPS}
\eea
which like instantons have a finite action 
${\cal S}=\f{-8\pi^2k}{g^2}$, $k<0$, with $k$ the instanton number. 
Since $k<0$, the index theorem implies that $\ola$ cannot be zero, and for   
the first equation above to be consistent with Eq. ({\ref{F1}) we must have $\la=0$. 
The solutions to the above equations, with $\la=0$, now solve the field equations (\ref{F1}) and (\ref{F2}). In fact we will see that a $C$ dependent contribution to 
the chiral gluino correlators arises exactly because of the existence of these classical field configurations. Below, we will examine the possible contributions of the above classical field configurations on both chiral and antichiral gluino correlators.  
Notice that both configurations above, i.e., anti-instantons and those of (\ref{BPS}), 
are supplemented with the ordinary Dirac equation for chiral and antichiral fermions.  
As discussed in \cite{SOH}, these are also the corresponding equations for the 
fermionic zero modes. Therefore, in a perturbative expansion around these solutions, 
one needs to take care of the zero modes of the Dirac operator by inserting  
appropriate operators in the path integral, just as in ${\cal N}=1$ SYM theory.

\section{Antichiral gluino condensates}

Let us begin with assuming that the superspace coordinates $\th^\al$ 
are not anticommuting, and instead they satisfy the following anticommutation 
relation
\be
\{\th^\al , \th^\bet \} = C^{\al\bet}\, , \label{DEF}
\ee
where $C^{\al\bet}$ is a constant and symmetric $2\times 2$ matrix. 
This deformation of the superspace has been studied earlier in \cite{SCH, FERR, KLEMM}.  
The anticommutation relation (\ref{DEF}) will deform the supersymmetry algebra with ${\overline Q}^2$ 
proportional to the deformation parameter $C^{\al\bet}$.
Seiberg \cite{SEI} has considered the above deformation in  
${\cal N} =1$ supersymmetric model and has shown that half of the 
supersymmetries can be preserved. 
Indeed if $W^\al = (A_\mu , \la)$ denotes the usual ${\cal N}=1$ gauge super 
multiplet, then the Lagrangian of this ${\cal N}=1/2$ model 
 reads
\bea 
{\cal L} &=& \f{i\tau}{16\pi} \int d^2\th \tr W^\al W_\al -\f{i{\bar \tau}}{16\pi}
\int d^2\thb 
\tr \ow^\alb\ow_\alb \\ \nn
&& + \f{(i\tau -i{\bar \tau})}{16\pi}\lf(-iC^{\mu\nu}\tr F_{\mu\nu}\ola\ola +\f{|C|^2}{4} 
\tr (\ola\ola)^2\ri)\, ,
\eea 
where
\[
C^{\mu\nu} \equiv C^{\al\bet}\ep_{\bet\ga}\si^{\mu\nu\ \ga}_\al \, 
\]
is a constant and antisymmetric self-dual matrix, with $|C|^2=C_{\mu\nu}C^{\mu\nu}$, 
and $\tau$ is the complex coupling constant
\be
\tau = \f{\th}{2\pi} + \f{4\pi i}{g^2}\, .
\ee
The above Lagrangian is invariant under the following $Q$ deformed supersymmetry 
transformations, 
\bea
&& \del \la = i\ep D +\si^{\mu\nu}\ep \lf(F_{\mu\nu}+\f{i}{2}C_{\mu\nu}\ola\ola\ri)\\ \nn 
&& \del A_\mu =-i\ola\osi_\mu \ep \\ \nn  
&& \del D= -\ep\si^\mu D_\mu\ola \\ \nn
&& \del\ola =0  \nn \, ,
\eea
whereas ${\overline Q}$ is broken. Solving for the auxillary field $D$, the 
action reads
\be
{\cal S} = \f{1}{2g^2}\int d^4 x\ \tr\! \lf ( \f{1}{2} F^{\mu\nu}F_{\mu\nu} +2i \la\dir \ola 
+i C^{\mu\nu}F_{\mu\nu}\ola\ola -\f{1}{4}\, |C|^2\, (\ola\ola)^2 \ri )\, .
\label{AC}
\ee

An interesting observation is that while the terms proportional to the deformation 
parameter $C^{\mu\nu}$ cannot be written as a $Q$ commutator, the variation of the Lagrangian with respect to $C^{\mu\nu}$ can:
\bea
\del {\cal L} &=& \f{i}{2g^2}\del C^{\mu\nu}\tr \lf( F_{\mu\nu}\ola\ola +\f{i}{2} 
C_{\mu\nu}\, (\ola\ola)^2 \ri)  \nn \\
&=& -\f{ i}{4g^2}\lf\{Q^\al\, ,\, \del C^{\mu\nu} 
(\si_{\mu\nu})_{\al\bet}\, \tr (\la^\bet \ola\ola) \ri \}\, . \label{LAG}
\eea
Assuming supersymmetry is not spontaneously broken, i.e., $Q\, |0\ran =0$, 
this will have an important consequence that the partition function does not 
depend on $C$:
\be
\f{\del Z}{\del C^{\mu\nu}}=- \lf\lan \f{\del {\cal S}}{\del C^{\mu\nu}} \ri\ran 
= \f{i}{4g^2}\int d^4x\lf\lan \lf\{Q^\al\, ,\,  
(\si_{\mu\nu})_{\al\bet}\, \tr(\la^\bet \ola\ola (x)) \ri \}\ri\ran = 0 \, ,
\ee
in other words, the effective action (obtained by doing the path integral 
over all fields) does not depend on the deformation parameter $C$. 
This even further extends to correlation functions of $Q$ invariant operators. 
Let ${\cal O}^i$, for $i=1,2,\ldots, n$, be operators such that 
$[ Q , {\cal O}^i\}=0$, then we will have
\bea
\f{\del}{\del C^{\mu\nu}} \lf\lan {\cal O}^1{\cal O}^2\ldots {\cal O}^n \ri\ran \!\! 
&=&\!\!\f{i}{4g^2}\int d^4x\lf\lan {\cal O}^1{\cal O}^2\ldots {\cal O}^n 
\lf\{Q^\al\, ,\,  
(\si_{\mu\nu})_{\al\bet}\, \tr(\la^\bet \ola\ola (x)) \ri \} \ri\ran \nn \\
&=&\!\!\f{i}{4g^2}\int d^4x\lf\lan  {\cal O}^1{\cal O}^2\ldots {\cal O}^n\,   
Q^\al\, (\si_{\mu\nu})_{\al\bet}\, \tr(\la^\bet \ola\ola (x)) \ri\ran \nn \\
&=&\!\!\pm \f{i}{4g^2}\int d^4x\lf\lan\, Q^\al {\cal O}^1{\cal O}^2\ldots {\cal O}^n  
(\si_{\mu\nu})_{\al\bet}\, \tr(\la^\bet \ola\ola) \ri\ran 
=0 \, . \label{Q}
\eea
Therefore, in computing the correlation functions of $Q$ invariant operators 
one can choose a convenient value of $C$. One such a choice is set $C$ to zero 
reducing the Lagrangian to that of ${\cal N}=1$ SYM theory. Hence, such operators 
will have the same expectation value as in pure ${\cal N}=1$ SYM theory. 
In particular, the antichiral gluino correlator will be invariant under the 
$C$ deformation:
\be
\f{\del}{\del C^{\mu\nu}}
\lan \tr\ola\ola(x_1)\, \tr\ola\ola(x_2)\ldots \tr\ola\ola(x_N) \ran =0 \, .
\ee
Moreover, one can use the Ward identity of the unbroken $Q$ supersymmetry to 
show that the above correlation function is $x$-independent. Cluster decomposition 
then implies that the operator $\tr (\ola\ola)$ gets a $C$ independent vacuum 
expectation value just as in pure ${\cal N}=1$ SYM theory \cite{AMATI}:
\be
\lan \tr (\ola\ola) \ran = A{\bar \Lambda}^3 \, ,
\ee
with the one-loop Renormalization Group (RG) invariant scale, ${\bar \Lambda}$, 
related to the microscopic cutoff ${ \Lambda}_0$ through
\be
{\bar \Lambda}^{3N} = \Lambda_0^{3N}\, e^{{-2\pi i\, {\overline \tau}}} \, .\label{L}
\ee

\section{Chiral sector}

What happens to chiral operators such as $\tr(\la\la)$? To see the consequence of 
$C$-deformation on the correlation functions of such operators we use a Ward 
identity corresponding to the $U(1)_R$ symmetry. To start with, let us first discuss 
the case of ${\cal N}=1$ SYM theory in some more detail. Classically there 
is a $U(1)_R$ symmetry acting on the fields as 
\be
\la \to e^{i\al}\la\, ,\ \ \ \ola \to e^{-i\al}\ola \, .\label{lala}
\ee
Quantum mechanically, however, $U(1)_R$ is reduced to $Z_{2N}$.  
Roughly speaking, this is  seen by noticng that in the background of  a gauge field 
with the topological charge $k$, the number of chiral and antichiral zero modes 
, $n_+$ and ${ n}_-$ respectively, are related to the index of the Dirac operator in that background as follows
\be
n_+ - {n}_-= {\rm index}\dir =2kN \, ,
\ee
and thus under $U(1)_R$ the measure of the path integral in this background transforms as
\be
{\cal D}\ola\, {\cal D}\la \to e^{-2iNk\al}\, {\cal D}\ola\, {\cal D}\la \, , \ \ \ \ \ \label{MEA}
\ee
whereas for nonzero modes the measure is invariant. Notice that in writing the measure 
we use the eigenfunctions of the Dirac operator as a basis, and that the above index theorem holds for any gauge field which has a topological charge $k$, independent of the equation it satisfies. 
Clearly ({\ref{MEA}) shows that the subgroup 
$Z_{2N}$ is left unbroken. This latter symmetry is further spontaneously broken 
to $Z_2$ by gluino condensations.  
Chiral gluino condensation happens because of the existence of anti-instantons with $k=1$. 
In this background there are $2N$ chiral zero modes, and to saturate them one inserts 
in the path integral $N$ gauge invariant operators made up of chiral fermions $\tr(\la\la)$ of the form 
\be
G(x_1,\ldots ,x_N)=\lan \tr\la\la(x_1)\, \tr\la\la(x_2)\ldots \tr\la\la(x_N) \ran   \, .
\label{CO}
\ee
In pure ${\cal N}=1$ SYM theory, one can also use the Ward identities of 
${\overline Q}$ 
symmetry (of the kind we used in eq. (\ref{Q})) to show that the above 
correlation function does not depend on $x_i$'s. Using this and cluster 
decomposition one then concludes that $\tr \la\la$ gets a vacuum expectation 
value, and hence a spontaneous breaking of $Z_{2N}$ to $Z_2$ leaving  
$N$ distinct quantum vacuua. Hence, we see a very similar behaviour between the chiral and antichiral sectors of ${\cal N}=1$ theory. 

The situation for ${\cal N}=1/2$ SYM theory is though quite different. 
In fact, since there is 
no ${\overline Q}$ supersymmetry, the whole argument above including the 
$x$-independence of $G(x_1,\ldots ,x_N)$, and the subsequent spontaneous breaking of $Z_{2N}$ to $Z_2$ breaks down. So the lack of ${\overline Q}$ supersymmetry 
forbids us from going any further in this direction. However, we can still 
use the anomalous $U(1)_R$ symmetry to discuss the possible $C$ dependence of 
correlation functions like (\ref{CO}). 
Before proceeding in the anomaly discussion note that, since anti-instantons ($F^-=0$) 
and the corresponding chiral zero modes (${\overline {\dir}}\la =0$) are  
solutions to the equations of motion, the operators inserted in (\ref{CO}) 
continue to saturate the $2N$ chiral zero modes of the Dirac operator in $k=1$ 
topological sector. 
Gauge fields of higher topological charge ($k>1$) give zero contribution 
because of the unsaturated zero modes ($n_+ = 2kN +n_-$). 
But, unlike pure ${\cal N}=1$ SYM theory, 
gauge fields with $k<0$ can also contribute to this correlation function. This happens, 
as we will see in the following, because the antichiral zero modes can be 
saturated by the $C$ dependent terms already present in the action. 

To obtain the Ward identity related to $U(1)_R$ symmetry, first notice that 
the terms proportional to $C$ in the action explicitly break the $U(1)_R$ symmetry: 
\be
\del_R {\cal S} =\f{\al}{2g^2} \int d^4x\, \tr\!\lf( 2 C^{\mu\nu}F_{\mu\nu}\ola\ola +i\, 
|C|^2\, (\ola\ola)^2 \ri )\, ,
\ee
for $\al$ an infinitesimal parameter of $U(1)_R$. On the other hand, in the 
background of gauge fields with topological charge $k$, which we expand around, 
according to (\ref{MEA}) the measure has a charge $-2kN$ under the $U(1)_R$ symmetry. 
Suppose now that an operator ${\cal O}$ with an $R$ charge equal to $q$ is inserted 
in the path integral. The invariance of the whole path integral, with the operators 
inserted in, under a change of field variables results to the following Ward identity
\be 
\lan {\cal O}\, \del_R {\cal S}\ran = i\al\,  (q-2Nk) \lan {\cal O}\ran \label{WARD}\, .
\ee
The above identity can be written in a more useful way. Using (\ref{LAG}), we have
\be
\del_R {\cal S} = -2i\al\, C^{\mu\nu}\f{\del {\cal S}}{\del C^{\mu\nu}}\, ,
\ee
so the Ward identity (\ref{WARD}) reads
\be
C^{\mu\nu}\f{\del }{\del C^{\mu\nu}}\lan {\cal O}\ran_k =\f{1}{2}
(q-2Nk) \lan {\cal O}\ran_k \, ,\label{WARD2}
\ee
where we have now put the subscript $k$ to indicate the corresponding topological sector. 
Let us first discuss the case with $k>0$. Taking ${\cal O}$ the operator inserted in (\ref{CO}) with $q=2N$, and choosing $k=1$, for instance, Ward identity (\ref{WARD}) becomes 
\be
\lf\lan \tr\la\la(x_1)\, \tr\la\la(x_2)\ldots \tr\la\la(x_N)\, \del_R {\cal S} 
\ri\ran_{k=1} =0\, ,
\ee
or equivalently
\be
\f{\del }{\del C^{}}
\lf\lan \tr\la\la(x_1)\, \tr\la\la(x_2)\ldots \tr\la\la(x_N) 
\ri\ran_{k=1} =0\, ,
\ee
where we have set
\be
C_{12}=C_{34} \equiv \f{C}{4} \, .\label{CC}
\ee
So for $k=1$ we can again take the limit where $C$ vanishes. This will reduce 
the computation to that of ${\cal N}=1$ theory, where we know only $k=1$ 
contributes to (\ref{CO}) \cite{AMATI}. 
Moreover, as said before, gauge fields of higher topological charge, $k>1$, do not 
contribute to (\ref{CO}), hence we will have
\be
\sum_{k\geq 1}\, G_k(x_1,\ldots ,x_N)
\, e^{2\pi ik\,  \tau}  = \lf\lan \tr\la\la(x_1)\, \tr\la\la(x_2)
\ldots \tr\la\la(x_N) \ri\ran_{C=0} = A\Lambda^{3N} \, .\label{AM}
\ee 

For $k\leq 0$, and $q=2N$, Ward identity (\ref{WARD2}) reads
\be   
C^{}\f{\del }{\del C^{}}\, G_k(x_1,\ldots ,x_N)
=(N+N|k|)\, G_k(x_1,\ldots ,x_N)\, \ \ \ \ \ k\leq 0\, ,
\ee
which, upon integration, results to the following $C$ dependence of (\ref{CO}) 
\be
G_k(x_1,\ldots ,x_N)\equiv \lf\lan \tr\la\la(x_1)\, \tr\la\la(x_2)\ldots \tr\la\la(x_N) 
\ri\ran_k\, = A_k\, C^{N+N|k|}\, ,\ \ \ \ k\leq 0 \, ,\label{CDEP}
\ee
where $A_k$ is in general a function of the coupling $g^2$, and the ultraviolet 
cutoff $\Lambda_0$. Also note that, as there is no ${\overline Q}$ symmetry, $A_k$ might well depend on $|x_i-x_j|$. A question that may now arise is what are the classical field 
configurations which give rise to such a $C$ dependent contributions?  
To answer this question let us look back at Eqs. (\ref{BPS}) and consider 
a perturbative calculation around the corresponding solutions. As discussed in Introduction, in this background there are no chiral zero modes, 
and according to the index theorem there are
$2|k|N$ antichiral zero modes of the Dirac operator. 
To saturate these zero modes, we need to pull down $2|k|N$ terms proportional to 
$\ola$ from the action.  
Moreover, as there are no chiral zero modes, $2N$ chiral fermions in the correlator 
must get Wick contracted with additional $2N$ antichiral fermions brought down from 
the action. Noticing that each factor of $\ola\ola$ comes with one power of $C$,  
then explains the factor of $C^{N+N|k|}$ appeared in (\ref{CDEP}). 
For the case of $U(2)$ gauge group, and as a schematic illustration of the above perturbative calculation, in Figures 1-4 we have depicted the related Feynman 
graphs to the lowest order in $g^2$. The subscript $0$ there indicates the zero modes 
(fermionic classical field configurations of (\ref{BPS})).

To sum up the contributions from all topological sectors characterized by the 
instanton number $k$, we still need to take into account the $\th$ angle 
contribution by multiplying each term in ({\ref{CDEP}) by a factor of  
$e^{i\th k}$. Moreover, since ${\cal S}=\f{-8\pi^2k}{g^2}$ in the background 
of (\ref{BPS}), the weight factor adds up to $e^{2\pi ik{\overline \tau}}$.  
After all, we arrive at 
\bea
G(x_1,\ldots ,x_N) &=& A\Lambda^{3N} + \sum_{k\leq 0}\, G_k(x_1,\ldots ,x_N)
\, e^{2\pi ik\, {\overline \tau}} \nn \\ 
&=& A\Lambda_0^{3N}\, e^{2\pi i \tau} + \sum_{k\leq 0} A_k\,  C^{N+N|k|}\, e^{2\pi ik\, {\overline \tau}}  \, ,\label{GLO}
\eea
where we have taken into account the term 
$A\Lambda^{3N}= A\Lambda_0^{3N}\, e^{2\pi i \tau}$ in (\ref{AM}). 
Notice that the above expression for $G(x_1,\ldots ,x_N)$ 
has the right $R$ charge. In fact if we regard $\tau$ $({\overline \tau})$ 
and $C$ as 
background fields and let them transform under the $U(1)_R$ group as follows 
\bea
\tau \to \tau + \f{\al N}{\pi} \nn \\
{\overline \tau} \to {\overline \tau} + \f{\al N}{\pi} \nn \\
C \to e^{2 i\al}\, C \, ,
\eea
then by looking at (\ref{lala}), (\ref{MEA}), and the action (\ref{AC}) 
we can see that the full path integral (either in the background of 
anti-instantons or generalized instantons (\ref{BPS})) 
is invariant under the $R$-symmetry group. Therefore, the expectation value of any 
operator must transform under the $R$-symmetry group exactly as the operator itself does. 
This is the behavior we observe in (\ref{GLO}).

Two further comments about our result in (\ref{GLO}) are in order. First note that the 
holomorphicity in $\tau$ of the chiral gluino correlators observed in ${\cal N}=1$ 
theory, is now spoilt by the appearance of ${\overline \tau}$ in $C$ dependent terms. 
The second comment concerns the calculation of $A_k$'s in the above correlation 
function. Upon taking the $N+N|k|$ derivative of (\ref{CDEP}) with respect to $C$, 
we have 
\be
\f{1}{(N+N|k|)!}\lf(\f{\del}{\del C}\ri)^{N+N|k|} G_k(x_1,\ldots ,x_N) = A_k \, .
\ee
As the right hand side is independent of $C$, we can take the limit of 
vanishing $C$ in the left hand side to reduce it to a specific correlation function 
in ${\cal N}=1$ SYM theory. In this way the computation of $A_k$'s 
can be reduced to a computation in ${\cal N}=1$ SYM theory.

\begin{center} 
\begin{picture}(300,100)(0,0)
\Oval(150,50)(20,47.2)(0)
\Gluon(150,69.5)(150,30.5){3}{5}
\Vertex(150,69.5){1.5}
\Vertex(150,30.5){1.5}
\Text(70,50)[]{$\tr(\la\la)(x_1)$}
\Text(230,50)[]{$\tr(\la\la)(x_2)$}
\Text(103,50)[]{$\times $}
\Text(197.4,50)[]{$\times $}
\Text(142,78)[]{$\ola$}
\Text(159,78)[]{$\ola$}
\Text(142,22)[]{$\ola$}
\Text(159,22)[]{$\ola$}
\Text(162,50)[]{$A_\mu$}
\end{picture}  \\ 
{\bf Figure 1.}{\sl\ 2-loop diagram in $k=0$ sector proportional to $C^2$} 
\end{center}
 
\begin{center} 
\begin{picture}(300,100)(0,0)
\Curve{(103,50)(128,70)(150,50)(175,30)(197.4,50)}
\Curve{(103,50)(128,30)(150,50)(175,70)(197.4,50)}
\Vertex(150,50){1.5}
\Text(70,50)[]{$\tr(\la\la)(x_1)$}
\Text(230,50)[]{$\tr(\la\la)(x_2)$}
\Text(103,50)[]{$\times $}
\Text(197.4,50)[]{$\times $}
\Text(150,65)[]{$\ola^4$}
\end{picture}  \\ 
{\bf Figure 2.}{\sl\ 2-loop diagram in $k=0$ sector proportional to $C^2$} 
\end{center}
\begin{center}
\begin{picture}(300,100)(0,0)
\Oval(120,50)(20,50)(0)
\Oval(180,50)(20,50)(0)
\Gluon(100,68)(100,32){3}{5}
\Gluon(200,68)(200,32){-3}{5}
\Text(90,77)[]{$\ola$}
\Text(110,78)[]{$\ola_0$}
\Text(90,24)[]{$\ola$}
\Text(110,22)[]{$\ola_0$}
\Text(112,50)[]{$A_\mu$}
\Text(212,50)[]{$A_\mu$}
\Text(190,78)[]{$\ola_0$}
\Text(208,77)[]{$\ola$}
\Text(193,22)[]{$\ola_0$}
\Text(209,24)[]{$\ola$}
\BCirc(150,50){25}
\PText(150,50)(0)[]{INST}
\Text(39,50)[]{$\tr(\la\la)(x_1)$}
\Text(262,50)[]{$\tr(\la\la)(x_2)$}
\Text(70.5,50)[]{$\times $}
\Text(230.5,50)[]{$\times $}
\Vertex(100,68){1.5}
\Vertex(100,32){1.5}
\Vertex(200,68){1.5}
\Vertex(200,32){1.5}
\end{picture}  \\ 
{\bf Figure 3.}{\sl\ 2-loop diagram in $k=-1$ sector proportional to $C^4$} 
\end{center}

\begin{center} 
\begin{picture}(300,100)(0,0)
\Curve{(54.6,50)(57,53)(70,63)(75,65)(103,50)(130,35)}
\Curve{(54.6,50)(57,47)(70,37)(75,35)(103,50)(130,65)}
\Curve{(170,35)(197.4,50)(226.3,65)(231.3,63)(244.3,53)(246.7,50)}
\Curve{(170,65)(197.4,50)(226.3,35)(231.3,37)(244.3,47)(246.7,50)}
\BCirc(150,50){25}
\PText(150,50)(0)[]{INST}
\Vertex(103,50){1.5}
\Vertex(197.4,50){1.5}
\Text(22,50)[]{$\tr(\la\la)(x_1)$}
\Text(280,50)[]{$\tr(\la\la)(x_2)$}
\Text(54.6,50)[]{$\times $}
\Text(246.7,50)[]{$\times $}
\Text(110,65)[]{$\ola_0$}
\Text(90,69)[]{$\ola$}
\Text(110,35)[]{$\ola_0$}
\Text(92,33)[]{$\ola$}
\Text(190,65)[]{$\ola_0$}
\Text(208,67)[]{$\ola$}
\Text(193,35)[]{$\ola_0$}
\Text(209,33)[]{$\ola$}
\end{picture}  \\ 
{\bf Figure 4.}{\sl\ 2-loop diagram in $k=-1$ sector proportional to $C^4$} 
\end{center}

\hspace{30mm}

\hspace{-6mm}{\large \textbf{Acknowledgement}}
\newline
It is a pleasure to thank R. Abbaspur, H. Arfaei, S. Parvizi, and S.-J. Rey for 
useful discussions.

\end{document}